\documentclass[print,superscriptaddress,showpacs,amsmath, amssymb,aps,pra]{revtex4-1}
\usepackage[T1]{fontenc}
\usepackage{amssymb}
\usepackage{graphicx}
\usepackage{dcolumn}
\usepackage{dsfont}
\usepackage{bm}
\usepackage{subfigure}
\usepackage[ruled]{algorithm2e}
\usepackage{hyperref}
\usepackage{appendix}

\hypersetup{colorlinks=true, citecolor=blue, urlcolor=blue, linkcolor=blue}
\begin{document}

\title{Tighter Constraints of Multipartite Systems in terms of General Quantum Correlations}

\author{Jin-Hong Hao}
\affiliation{School of Mathematical Sciences, Capital Normal University, Beijing 100048, China}
\author{Ya-Ya Ren}
\affiliation{School of Mathematical Sciences, Capital Normal University, Beijing 100048, China}
\author{Qiao-Qiao Lv}
\affiliation{School of Mathematical Sciences, Capital Normal University, Beijing 100048, China}
\author{Zhi-Xi Wang}
\affiliation{School of Mathematical Sciences, Capital Normal University, Beijing 100048, China}
\author{Shao-Ming Fei}
\affiliation{School of Mathematical Sciences, Capital Normal University, Beijing 100048, China}
\affiliation{Max-Planck-Institute for Mathematics in the Sciences, 04103, Leipzig, Germany}

\begin{abstract}
\centerline {\textbf{Abstract}}
Monogamy and polygamy relations characterize the quantum correlation distributions among multipartite quantum systems.
We investigate the monogamy and polygamy relations satisfied by measures of general quantum correlation.
By using the Hamming weight, we derive new monogamy and polygamy inequalities
satisfied by the $\beta$-th power and the $\alpha$-th power of general quantum correlations, respectively. We show that
these monogamy and polygamy relations are tighter than the existing ones, such as [Int. J. Theor. Phys. 60, 1455-1470 (2021)]. Taking concurrence and the Tsallis-$q$ entanglement of assistance as examples, we show the advantages of our results.

\end{abstract}

\maketitle

\section{INTRODUCTION}
Quantum entanglement \cite{MAN,RPMK} is an essential feature of quantum mechanics. It distinguishes quantum mechanics from the classical world. They also play an important role in quantum teleportation \cite{bbc}, nosignaling theories \cite{AGMD}, and quantum computation \cite{rh,JM,JML}. Entanglement is also important in quantum key distribution \cite{ek}, quantum secure direct communication \cite{LGL}, testing effective field theories \cite{YJH}, quantum steering \cite{XS}. At present, the entanglement in a 53 IBM quantum computing system has been studied \cite{HWJ}. To quantify the quantum entanglement \cite{VMMP}, many entanglement measures have been proposed such as entanglement of formation \cite{W}, concurrence \cite{SW}, negativity \cite{TMG}, and so on. One of the most important issues closely related to the entanglement measure is the monogamy and polygamy relations satisfied by the entanglement \cite{B}. Monogamy and polygamy relations characterize the distribution \cite{MMV} of the entanglement in multipartite systems. Polygamy relation is important in high dimensional system \cite{JRomero}.
	
For a three-qubit quantum state $\rho_{ABC}$ the monogamy relation \cite{CKW} is characterized by $\mathcal{E}(\rho_{A|BC})\geq \mathcal{E}(\rho_{AB}) +\mathcal{E}(\rho_{AC})$, where $\mathcal{E}$ is a bipartite entanglement measure, $\mathcal{E}(\rho_{A|BC})$ is the entanglement under the bipartition $A$ and $BC$, $\rho_{AB}$ and $\rho_{AC}$ are the reduced density matrices of $\rho_{ABC}$. However, such monogamy relations depend on detailed entanglement measures and quantum states. It has been proved that the squared concurrence $C^2$ \cite{TJ,YKM} and the entanglement of formation $E^2$ \cite{TR} satisfy the monogamy relations for multiqubit states. Later it has been generalized to various entanglement measures, such as continuous-variable entanglement \cite{agf}, Tsallis-$q$ entropy \cite{kjs,kjsg}, Renyi-entropy entanglement \cite{wmv}, and so on. Dual to the monogamy relation, the polygamy relation \cite{gg} is described by $\mathcal{E}(\rho_{A|BC})\leq \mathcal{E}(\rho_{AB}) +\mathcal{E}(\rho_{AC})$. Polygamy relations have been similarly studied under different entanglement measures \cite{JQ} for multipartite quantum systems as well as some higher-dimensional quantum systems  \cite{qjin,fjin,Kim}.

Monogamy and polygamy relations characterize the distributions of quantum correlations in multipartite systems. Tighter monogamy relations give rise to better characterization of the quantum correlation distributions \cite{ggt}. In this paper, we investigate the monogamy and polygamy inequalities for general measures of quantum correlations in multipartite systems. These inequalities are proved to be tighter than the existing ones. Taking the concurrence and Tsallis-$q$ entropy as examples, we show the advantages of our general monogamy and polygamy inequalities.

\section{TIGHTER MONOGAMY RELATIONS FOR GENERAL QUANTUM CORRELATIONS}

We denote $\mathcal{Q}$ a measure of general quantum correlation for bipartite systems. A monogamy inequality
for an $N$-partite quantum state $\rho_{AB_1B_2,\cdots,B_{N-1}}$ has the form \cite{ARA},
\begin{eqnarray}\label{q}
\mathcal{Q}(\rho_{A|B_1B_2,\cdots,B_{N-1}})\geq\mathcal{Q}(\rho_{AB_1})+\mathcal{Q}(\rho_{AB_2})+\cdots+\mathcal{Q}(\rho_{AB_{N-1}}),
\end{eqnarray}
where $\rho_{AB_i}$, $i=1,...,N-1$, are the reduced density matrices. For simplicity, we denote $\mathcal{Q}(\rho_{AB_i})$ by $\mathcal{Q}_{AB_i}$, and $\mathcal{Q}(\rho_{A|B_1B_2,\cdots,B_{N-1}})$ by $\mathcal{Q}_{A|B_1B_2,\cdots,B_{N-1}}$.
Generally, (\ref{q}) is not satisfied for arbitrary $\mathcal{Q}$. It has been proved that for arbitrary dimensional tripartite states, there exists a real number $\gamma$ $(\gamma\geq1)$ such that any quantum correlation measure $\mathcal{Q}$ satisfies the following monogamy relation \cite{sjin,SPAU}, $\mathcal{Q}^x_{A|BC}\geq\mathcal{Q}^x_{AB}+\mathcal{Q}^x_{AC}$, for $x\geq \gamma$. By using the inequality $(1+t)^x\geq 1+t^x$ for $x\geq1$, $0\leq t\leq1$, it is easy to generalize the result to the $N$-partite case,
\begin{eqnarray}\label{mq}
\mathcal{Q}^\gamma_{A|B_0B_1,\cdots,B_{N-1}}\geq \sum_{i=0}^{N-1}\mathcal{Q}_{AB_i}^\gamma,
\end{eqnarray} where $i=0,1,\cdots, N-1$.

 Firstly, we introduce two inequalities. For any real numbers $k$ ($0<k\leq 1$), $\delta$ ($\delta\geq 1$), $t$ ($0\leq t\leq k^\delta$) and non-negative real number $x$, $y$, one has \cite{LYY},
\begin{eqnarray}\label{ab}
(1+t)^x \geq 1+\mathcal{K}^\delta_{x}t^x, ~~ for ~~ x \geq 1,
\end{eqnarray}
\vspace{-0.7cm}
\begin{eqnarray}\label{ab1}
\hspace{0.8cm}(1+t)^y \leq  1+\mathcal{K}^\delta_{y}t^y, \quad for \quad 0 \leq y \leq 1,
\end{eqnarray}
where $\mathcal{K}^\delta_ x=\frac{(1+k^\delta)^x-1}{k^{\delta x}}$ and $\mathcal{K}^\delta_ y=\frac{(1+k^\delta)^y-1}{k^{\delta y}}$ .

We next present general monogamy inequalities for multipartite systems, satisfied by the $\alpha$-th power of any quantum correlation measures
in terms of the Hamming weight of the binary vectors related to the distribution of subsystems.
For any non-negative integer $j$ and its binary expansion
$j=\sum_{i=0}^{n-1}j_i2^i$, with $\log_2j\leq n$ and $j_i\in \{0,~1\}$, $i=0,1,\cdots,n-1$, one has the unique binary vector $\vec{j}$ associated with $j$,
$\vec{j}=(j_0,j_1,\cdots,j_{n-1})$. The Hamming weight $w_H(\vec{j})$ of $\vec{j}$ is defined to be the number of $1's$ in $\{j_0,j_1,\cdots,j_{n-1}\}$ \cite{MAN}.

[\emph{Theorem 1}] For any $(N+1)$-partite state $\rho_{AB_0\cdots B_{N-1}}$, if $k^\delta\mathcal{Q}_{AB_j}\geq \mathcal{Q}_{AB_{j+1}}\geq 0$ for $j=0, 1, \cdots,  N-2$, then for any real numbers $\alpha\geq\gamma$, $0<k^\delta\leq 1$ and $\delta\geq 1$, we have
\begin{eqnarray}\label{th72}
\mathcal{Q}^\alpha_{A|B_0B_1\cdots B_{N-1}}\geq \sum_{j=0}^{N-1} \mathcal{K}^{\delta {\omega_H(\vec{j})}}_\alpha\mathcal{Q}^\alpha_{AB_j},
\end{eqnarray}
where  $\mathcal{K}^\delta_\alpha=\frac{(1+k^\delta)^\frac{\alpha}{\gamma}-1}{k^{\delta \frac{\alpha}{\gamma}}}$.

[\emph{Proof}] According to (\ref{mq}), it is adequate to prove that
\begin{eqnarray}\label{pfth11}
\left(\sum_{j=0}^{N-1} \mathcal{Q}_{AB_j}^\gamma\right)^\frac{\alpha}{\gamma}\geq \sum_{j=0}^{N-1} \mathcal{K}^{\delta \omega_H(\vec{j})}_\alpha\mathcal{Q}^\alpha_{AB_j}.
\end{eqnarray}

We first show that the inequality (\ref{pfth11}) holds when $N = 2^s$ by using mathematical induction on $s$.
Let $\rho_{AB_0}$ and $\rho_{AB_1}$ be the reduced density matrices of a three-partite state $\rho_{AB_0B_1}$. For $s=1$, we have
\begin{eqnarray*}\label{pfth12}
\mathcal{Q}^\alpha_{A|B_0B_1}&\geq\left(\mathcal{Q}^\gamma_{AB_0}+ \mathcal{Q}^\gamma_{AB_1}\right)^\frac{\alpha}{\gamma}=\mathcal{Q}^\alpha_{AB_0} \left(1+\frac{\mathcal{Q}^\gamma_{AB_1}}{\mathcal{Q}^\gamma_{AB_0}}\right)^\frac{\alpha}{\gamma}
\geq\mathcal{Q}^\alpha_{AB_0} \left[1+\mathcal{K}^\delta_\alpha\left(\frac{\mathcal{Q}_{AB_1}}{{\mathcal{Q}}_{AB_0}}\right)^{\alpha}\right]
=\mathcal{Q}^\alpha_{AB_0}+\mathcal{K}^\delta_\alpha\mathcal{Q}^\alpha_{AB_1},
\end{eqnarray*}
where the first inequality is due to that the function $f(x)=x^t$ is increasing in $(0,+\infty)$ when $t>0$, the second inequality is due to the inequality (\ref{ab}).
Now we assume that the inequality (\ref{pfth11}) is true for $N=2^{s-1}$ for $s \geq 2$. Consider $N=2^s$. For $(N+1)$-partite state $\rho_{AB_0\cdots B_{N-1}}$ we have
\begin{eqnarray}
\left(\sum_{j=0}^{N-1} \mathcal{Q}_{AB_j}^\gamma\right)^\frac{\alpha}{\gamma}
&=&\left(\sum_{j=0}^{2^{s-1}-1} \mathcal{Q}_{AB_j}^\gamma+\sum_{j=2^{s-1}}^{2^s-1} \mathcal{Q}_{AB_j}^\gamma\right)^\frac{\alpha}{\gamma}\nonumber\\
&=&\left(\sum_{j=0}^{2^{s-1}-1} \mathcal{Q}_{AB_j}^\gamma\right)^\frac{\alpha}{\gamma}\left(1+\frac{\sum_{j=2^{s-1}}^{2^s-1} \mathcal{Q}_{AB_j}^\gamma}{\sum_{j=0}^{2^{s-1}-1} \mathcal{Q}_{AB_j}^\gamma}\right)^\frac{\alpha}{\gamma}\nonumber.
\end{eqnarray}
In light of the conditions of the theorem, we obtain
\begin{eqnarray}\nonumber
\sum_{j=2^{s-1}}^{2^s-1} \mathcal{Q}_{AB_j}^\gamma\leq k^\delta\sum_{j=0}^{2^{s-1}-1} \mathcal{Q}_{AB_j}^\gamma.
\end{eqnarray}
Thus, we have
\begin{eqnarray}\nonumber
\left(\sum_{j=0}^{N-1}\mathcal{Q}_{AB_j}^\gamma\right)^\frac{\alpha}{\gamma}
\geq\left(\sum_{j=0}^{2^{s-1}-1} \mathcal{Q}_{AB_j}^\gamma\right)^\frac{\alpha}{\gamma}+\mathcal{K}^\delta_\alpha\left(\sum_{j=2^{s-1}}^{2^s-1} \mathcal{Q}_{AB_j}^\gamma\right)^\frac{\alpha}{\gamma}.
\end{eqnarray}
Using the induction hypothesis, we can easily get
\begin{eqnarray}\label{pfth15}
\left(\sum_{j=0}^{2^{s-1}-1}\mathcal{Q}_{AB_j}^\gamma\right)^\frac{\alpha}{\gamma}\geq \sum_{j=0}^{2^{s-1}-1} \mathcal{K}^{\delta \omega_H(\vec{j})}_\alpha \mathcal{Q}_{AB_j}^\alpha,
\end{eqnarray}
and
\begin{eqnarray}\label{pfth16}
\left(\sum_{j=2^{s-1}}^{2^{s}-1}\mathcal{Q}_{AB_j}^\gamma\right)^\frac{\alpha}{\gamma} \geq \sum_{j=2^{s-1}}^{2^{s}-1} \mathcal{K}^{\delta \left(\omega_H(\vec{j})-1\right)}_\alpha\mathcal{Q}_{AB_j}^\alpha.
\end{eqnarray}
Combining the inequalities (\ref{pfth15}) and (\ref{pfth16}), we obtain
\begin{eqnarray}\label{pfth17}
\left(\sum_{j=0}^{2^{s}-1}\mathcal{Q}_{AB_j}^\gamma\right)^\frac{\alpha}{\gamma}\geq\sum_{j=0}^{2^{s}-1} \mathcal{K}^{\delta \omega_H(\vec{j})}_\alpha\mathcal{Q}_{AB_j}^\alpha.
\end{eqnarray}

For arbitrary positive integer $N$, without loss of generality, we assume that $0 < N\leq 2^s$ for some $s$. Considering a $(2^s+1)$-partite state
\begin{eqnarray}\label{pfth18}
\rho'_{AB_0\cdots B_{2^s-1}}=\rho_{AB_0\cdots B_{N-1}}\otimes\eta_{B_N\cdots B_{2^s-1}},
\end{eqnarray}
which is a product of $\rho_{AB_0\cdots B_{N-1}}$ and an arbitrary $(2^s-N )$-partite state $\eta_{B_N\cdots B_{2^s-1}}$.

We have
\begin{eqnarray}\label{pfth19}
\mathcal{Q}^\alpha(\rho'_{A|B_0B_1\cdots B_{2^s-1}})\geq\sum_{j=0}^{2^s-1} \mathcal{K}^{\delta\omega_H(\vec{j})}_\alpha\mathcal{Q}^\alpha(\sigma_{AB_j}),
\end{eqnarray}
where $\sigma_{AB_j}$ is the bipartite reduced density matrix of $\rho'_{AB_0\cdots B_{2^s-1}}$, $j=0,1,\cdots,2^s-1$. Obviously,
$\mathcal{Q}(\rho'_{A|B_0B_1\cdots B_{2^s-1}})=\mathcal{Q}(\rho_{A|B_0B_1\cdots B_{N-1}})$, $\mathcal{Q}(\sigma_{AB_j})=0$ for $j=N,\cdots,2^s-1$,
and $\sigma_{AB_j}=\rho_{AB_j}$ for $j=0,1,\cdots,N-1$. Hence,
\begin{eqnarray}\label{pfth113}
\mathcal{Q}^\alpha(\rho_{A|B_0B_1\cdots B_{N-1}})=\mathcal{Q}^\alpha(\rho'_{A|B_0B_1\cdots B_{2^s-1}})\geq\sum_{j=0}^{N-1} \mathcal{K}^{\delta\omega_H(\vec{j})}_\alpha\mathcal{Q}^\alpha(\rho_{AB_j}).
\end{eqnarray}
\hfill$\Box$

Theorem 1 gives a tighter monogamy relation based on the Hamming weight for arbitrary $(N+1)$-partite states.
Since $\mathcal{K}^{\delta\omega_H(\vec{j})}_\alpha\geq 1$ for any $\alpha\geq\gamma$, we have
\begin{eqnarray}\label{pfth114}
\mathcal{Q}^\alpha_{A|B_0B_1\cdots B_{N-1}}\geq\sum_{j=0}^{N-1} \mathcal{K}^{\delta\omega_H(\vec{j})}_\alpha \mathcal{Q}^\alpha_{AB_j}
\geq \sum_{j=0}^{N-1} \mathcal{Q}^\alpha_{AB_j}.
\end{eqnarray}
Therefore, the inequality (\ref{th72}) in Theorem 1 is generally tighter than the inequality (\ref{mq}). Moreover, Theorem 1 includes the results in \cite{LD} as a special case.

[\emph{Corollary 1}] For any multipartite state $\rho_{AB_0\cdots B_{N-1}}$, if $k^\delta\mathcal{Q}^\gamma_{AB_i}\geq\sum_{j=i+1}^{N-1} \mathcal{Q}^\gamma_{AB_j}$ for $j=0,1,\cdots N-2$, then for any $\alpha\geq\gamma$, $0<k^\delta\leq 1$ and $\delta\geq 1$, we have
$\mathcal{Q}^\alpha_{A|B_0B_1\cdots B_{N-1}}\geq\sum_{j=0}^{N-1} \mathcal{K}^{\delta j}_\alpha\mathcal{Q}^\alpha_{AB_j}$
where $\mathcal{K}^\delta_\alpha=\frac{(1+k^\delta)^\frac{\alpha}{\gamma}-1}{k^{\delta\frac{\alpha}{\gamma}}}$.

[\emph{Proof}] From the inequalities (\ref{mq}) and (\ref{ab}), we have
\begin{eqnarray}
\mathcal{Q}^{\alpha}_{A|B_0B_1\cdots B_{N-1}}
&\geq&\mathcal{Q}^{\alpha}_{AB_0}+\mathcal{K}^\delta_\alpha \left(\sum_{j=1}^{N-1}\mathcal{Q}_{AB_j}^\gamma\right)^\frac{\alpha}{\gamma}\nonumber\\
&\geq&\mathcal{Q}^{\alpha}_{AB_0}+\mathcal{K}^\delta_\alpha \mathcal{Q}^{\alpha}_{AB_1}
+\mathcal{K}^\delta_\alpha \left(\sum_{j=2}^{N-1}\mathcal{Q}_{AB_j}^\gamma\right)^\frac{\alpha}{\gamma}\nonumber\\
&\geq&\cdots\nonumber\\
&\geq&\mathcal{Q}^{\alpha}_{AB_0}+\mathcal{K}^\delta_\alpha\mathcal{Q}^{\alpha}_{AB_1}+\cdots+\mathcal{K}^{\delta(N-1)}_\alpha
\mathcal{Q}^{\alpha}_{AB_{N-1}}\nonumber.
\end{eqnarray}   $\hfill\Box$

For any non-negative integer $j$ and its corresponding binary vector $\vec{j}$, the Hamming weight $\omega_H(\vec{j})$ satisfies  $\omega_H(\vec{j})\leq j$, which indicates to $\mathcal{Q}^\alpha_{A|B_0B_1\cdots B_{N-1}}\geq\sum_{j=0}^{N-1} \mathcal{K}^{\delta j}_\alpha\mathcal{Q}^\alpha_{AB_j}\geq \sum_{j=0}^{N-1} \mathcal{K}^{\delta\omega_H(\vec{j})}_\alpha\mathcal{Q}^\alpha_{AB_j}$,
for any $\alpha\geq \gamma$.

[{\it Example 1}] Let us consider the three-qubit state $|\psi\rangle_{ABC}$ in the generalized Schmidt decomposition form
\begin{eqnarray}\label{ex1}
|\psi\rangle_{ABC}=\lambda_0|000\rangle+\lambda_1e^{i{\varphi}}|100\rangle+\lambda_2|101\rangle+\lambda_3|110\rangle+\lambda_4|111\rangle,
\end{eqnarray}
where $\lambda_i\geq0$, $i=0, 1, 2, 3, 4$, $\sum\limits_{i=0}\limits^4\lambda_i^2=1.$

For a bipartite pure state $|\psi\rangle_{AB}$, the concurrence is given by $C(|\psi\rangle_{AB})=\sqrt{{2\left[1-\mathrm{Tr}(\rho_A^2)\right]}}$,
where $\rho_A$ is the reduced density matrix obtained by tracing over the subsystem $B$ \cite{SW}. For an $N$-qubit state $\rho_{AB_1\cdots B_{N-1}}$, the concurrence $C(\rho_{A|B_1\cdots B_{N-1}})$ of the state $\rho_{AB_1\cdots B_{N-1}}$, viewed as a bipartite state under the partition $A$ and $B_1, B_2, \cdots, B_{N-1}$, satisfies the Coffman-Kundu-Wootters inequality \cite{TJ},
\begin{eqnarray}\label{C2}
C^\alpha(\rho_{A|B_1,B_2\cdots,B_{N-1}})\geq \sum_{i=1}^{N-1}C^\alpha(\rho_{AB_i}),
\end{eqnarray}
for $\alpha\geq2$. It is further improved that for $\alpha\geq2$, if $C(\rho_{AB_i})\geq C(\rho_{A|B_{i+1}\cdots B_{N-1}})$ for $i=1, 2, \cdots, N-2$, $N\geq 4$, then
\begin{eqnarray}\label{mo2}
C^\alpha(\rho_{A|B_1B_2\cdots B_{N-1}})\geq C^\alpha(\rho_{AB_1})+\frac{\alpha}{2} C^\alpha(\rho_{AB_2})+\cdots+\left(\frac{\alpha}{2}\right)^{N-2}C^\alpha(\rho_{AB_{N-1}}).
\end{eqnarray}
Moreover, in \cite{LD}, the author provide a monogamy relation for concurrence,
\begin{eqnarray}\label{L2}
{C^\alpha}_{A|B_0B_1\cdots B_{N-1}}\geq \sum_{j=0}^{N-1}\mathcal{K}_\alpha^{\omega_H(\vec{j})}{C^\alpha}_{AB_j}.
\end{eqnarray}

From (\ref{th72}) and (\ref{C2}), we have a tighter monogamy relation for concurrence $(\gamma=2)$,
\begin{eqnarray}\label{n3}
{C^\alpha}_{A|B_0B_1\cdots B_{N-1}}\geq \sum_{j=0}^{N-1}\mathcal{K}_\alpha^{\delta\omega_H(\vec{j})}{C^\alpha}_{AB_j}.
\end{eqnarray}
Since $\mathcal{K}_\alpha^{\delta \omega_H(\vec{j})}\geq \mathcal{K}_\alpha^{\omega_H(\vec{j})} $, one has ${C^\alpha}_{A|B_0B_1\cdots B_{N-1}}\geq\sum_{j=0}^{N-1}\mathcal{K}_\alpha^{\delta\omega_H(\vec{j})}{C^\alpha}_{AB_j}\geq \sum_{j=0}^{N-1}\mathcal{K}_\alpha^{\omega_H(\vec{j})}{C^\alpha}_{AB_j} $.
Therefore, (\ref{n3}) is tighter than the inequality (\ref{L2}) given in \cite{LD}.

From the definition of concurrence, we have $C_{A|BC}=2\lambda_0\sqrt{{\lambda_2^2+\lambda_3^2+\lambda_4^2}}$, $C_{AB}=2\lambda_0\lambda_2$ and $C_{AC}=2\lambda_0\lambda_3$. Set $\lambda_0=\lambda_3=\frac{1}{2}$, $\lambda_2=\frac{\sqrt{2}}{2}$ and $\lambda_1=\lambda_4=0$.
One has $C_{A|BC}=\frac{\sqrt{3}}{2}$, $C_{AB}=\frac{\sqrt{2}}{2}$, $C_{AC}=\frac{1}{2}$, $y_0\equiv C^{\alpha}_{A|BC}=\left(\frac{\sqrt{3}}{2}\right)^\alpha$. Choosing
$k =\frac{9}{10}$ and $\delta=2$, we have the lower bound of (\ref{n3}),
\begin{eqnarray}\
y_1&\equiv &C^{\alpha}_{AB}+\mathcal{K}^{\delta}_\alpha C^{\alpha}_{AC}=\left(\frac{\sqrt{2}}{2}\right)^\alpha+\frac{(1+0.9^2)^\frac{\alpha}{2}-1}{(0.9^2)^\frac{\alpha}{2}}\left(\frac{1}{2}\right)^\alpha,
\end{eqnarray}
and the lower bound of (\ref{L2}),
\begin{eqnarray}\
y_2&\equiv &C^{\alpha}_{AB}+\mathcal{K}_\alpha C^{\alpha}_{AC}
=\left(\frac{\sqrt{2}}{2}\right)^\alpha+\frac{(1+0.9)^\frac{\alpha}{2}-1}{(0.9)^\frac{\alpha}{2}}\left(\frac{1}{2}\right)^\alpha.
\end{eqnarray}
Fig. 1 shows that our bound $y_1$ is tighter than the bound $y_2$ from \cite{LD}.
\begin{figure}
\centering\includegraphics[width=7cm]{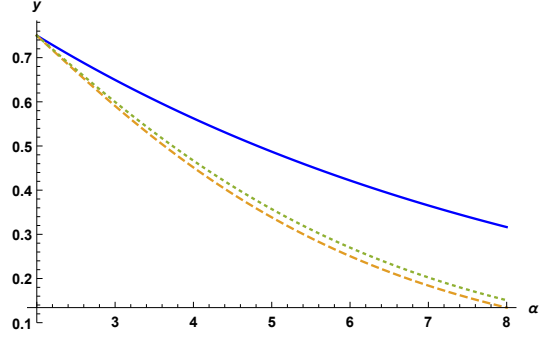}
\caption{The $y$ axis is the lower bound of ${C}^\alpha_{A|BC}$. The blue solid line for $y_0$, the green dotted line for $y_1$ and the yellow dashed line for $y_2$.}
\end{figure}

\section{TIGHTER POLYGAMY RELATIONS FOR GENERAL QUANTUM CORRELATIONS}

Next we provide a class of general polygamy inequalities in terms of the power of the quantum correlation measure $\mathcal{Q}$ and the Hamming weight of the binary vectors related to the distribution of subsystems. In \cite{sjin}, it has been proved that for any state $\rho_{AB_0\cdots B_{N-1}}$, there exists a real number $\gamma$ such that a quantum correlation measure $\mathcal{Q}$ satisfies the following polygamy relation, $\mathcal{Q}^y_{A|B_0\cdots B_{N-1}}\leq\sum_{j=0}^{N-1}\mathcal{Q}^y_{AB_j}$, for $0\leq y\leq \gamma$.

[\emph{Theorem 2}] For any $(N+1)$-partite quantum state $\rho_{AB_0\cdots B_{N-1}}$, if $k^\delta \mathcal{Q}_{AB_j}\geq \mathcal{Q}_{AB_{j+1}}\geq 0$
for $j=0,1,\cdots N-2$, then we have
\begin{eqnarray}\label{th92}
\mathcal{Q}^\beta_{A|B_0B_1\cdots B_{N-1}}\leq\sum_{j=0}^{N-1} \mathcal{K}^{\delta\omega_H(\vec{j})}_\beta{\mathcal{Q}^\beta}_{AB_j},
\end{eqnarray}
for $0\leq\beta\leq\gamma$, $0<k^\delta \leq 1$ and $\delta\geq 1$, where
$\mathcal{K}^\delta_\beta=\frac{(1+k^{\delta})^\frac{\beta}{\gamma}-1}{k^{\delta \frac{\beta}{\gamma}}}$.

[\emph{Proof}] It is adequate to show that
\begin{eqnarray}\label{pfth42}
\left(\sum_{j=0}^{N-1}\mathcal{Q}^\gamma_{AB_j}\right)^\frac{\beta}{\gamma}\leq\sum_{j=0}^{N-1}\mathcal{K}^{\delta \omega_H(\vec{j})}_\beta{\mathcal{Q}^\beta}_{AB_j}.
\end{eqnarray}

We first show that the inequality (\ref{pfth42}) holds when $N = 2^s$ by using mathematical induction on $s$.
For $s=1$ and a tripartite state $\rho_{AB_0B_1}$, from the inequality (\ref{ab1}) we have
\begin{equation*}
  (\mathcal{Q}^\gamma_{AB_0}+\mathcal{Q}^\gamma_{AB_1})^\frac{\beta}{\gamma}\leq  \mathcal{Q}^\beta_{AB_0}+\mathcal{K}^\delta_\beta \mathcal{Q}^\beta_{AB_1}.
\end{equation*}

Now we assume that the inequality (\ref{pfth42}) is true for $N=2^{s-1}$ with $s \geq 2$. Consider $(N+1)$-partite state $\rho_{AB_0\cdots B_{N-1}}$. We have for $N=2^s$,
\begin{eqnarray}\label{pfth43}
\left(\sum_{j=0}^{N-1} \mathcal{Q}^\gamma_{AB_j}\right)^\frac{\beta}{\gamma}= \left(\sum_{j=0}^{2^{s-1}-1} \mathcal{Q}^\gamma_{AB_j}\right)^\frac{\beta}{\gamma}\left(1+\frac{\sum_{j=2^{s-1}}^{2^s-1} \mathcal{Q}^\gamma_{AB_j}}{\sum_{j=0}^{2^{s-1}-1} \mathcal{Q}^\gamma_{AB_j}}\right)^\frac{\beta}{\gamma}.
\end{eqnarray}

From the conditions of the theorem, we have
\begin{eqnarray}\label{pfth44}
\sum_{j=2^{s-1}}^{2^s-1} \mathcal{Q}^\gamma_{AB_j}\leq k^\delta \sum_{j=0}^{2^{s-1}-1} \mathcal{Q}^\gamma_{AB_j}.
\end{eqnarray}
Again using the inequality (\ref{ab1}), we get
\begin{eqnarray}\label{pfth45}
\left(\sum_{j=0}^{N-1} \mathcal{Q}^\gamma_{AB_j}\right)^\frac{\beta}{\gamma} &&\leq\left(\sum_{j=0}^{2^{s-1}-1} \mathcal{Q}^\gamma_{AB_j}\right)^\frac{\beta}{\gamma}+\mathcal{K}^\delta_\beta\left(\sum_{j=2^{s-1}}^{2^s-1} \mathcal{Q}^\gamma_{AB_j}\right)^\frac{\beta}{\gamma}.
\end{eqnarray}
From the induction hypothesis, we can easily get that
\begin{eqnarray}\label{pfth46}
\left(\sum_{j=0}^{2^{s-1}-1} \mathcal{Q}^\gamma_{AB_j}\right)^\frac{\beta}{\gamma}\leq\sum_{j=0}^{2^{s-1}-1}\mathcal{K}^{\delta \omega_H(\vec{j})}_\beta\mathcal{Q}^\beta_{AB_j},
\end{eqnarray}
and
\begin{eqnarray}\label{pfth47}
\left(\sum_{j=2^{s-1}}^{2^{s}-1} \mathcal{Q}^\gamma_{AB_j}\right)^\frac{\beta}{\gamma}\leq\sum_{j=2^{s-1}}^{2^{s}-1} \mathcal{K}^{\delta (\omega_H(\vec{j})-1)}_\beta\mathcal{Q}^\beta_{AB_j}.
\end{eqnarray}
Thus, we have
\begin{eqnarray}\label{pfth48}
\left(\sum_{j=0}^{2^{s}-1} \mathcal{Q}^\gamma_{AB_j}\right)^\frac{\beta}{\gamma}\leq\sum_{j=0}^{2^{s}-1} \mathcal{K}^{\delta \omega_H(\vec{j})}_\beta\mathcal{Q}^\beta_{AB_j}.
\end{eqnarray}

Now consider $(N + 1)$-partite state $\rho_{AB_0\cdots B_{N-1}}$ for arbitrary positive integer $N$.
We can always assume that $0< N\leq 2^s$ for some $s$. For the $(2^s+1)$-partite state $\rho'_{AB_0\cdots B_{2^s-1}}$ (\ref{pfth18}),
we have
\begin{eqnarray}\label{pfth49}
\mathcal{Q}^\beta(\rho'_{A|B_0B_1\cdots B_{2^s-1}})\leq\sum_{j=0}^{2^s-1}\mathcal{K}^{\delta\omega_H(\vec{j})}_\beta\mathcal{Q}^\beta(\sigma_{AB_j}),
\end{eqnarray}
where $\sigma_{AB_j}$ is the bipartite reduced density matrix of $\rho'_{AB_0\cdots B_{2^n-1}}$, $j=0,1,\cdots,2^s-1$.
Moreover, since $\rho'_{AB_0B_1\cdots B_{2^s-1}}$ is a product state of $\rho_{AB_0\cdots B_{N-1}}$ and $\eta_{B_N\cdots B_{2^s-1}}$, we have
$\mathcal{Q}(\rho'_{A|B_0B_1\cdots B_{2^s-1}})=\mathcal{Q}(\rho_{A|B_0B_1\cdots B_{N-1}})$,
$\mathcal{Q}(\sigma_{AB_j})=0$ for $j=N,\cdots,2^s-1$, and $\sigma_{AB_j}=\rho_{AB_j}$ for $j=0,1,\cdots,N-1$.
Therefore, we get
\begin{eqnarray}\nonumber
\mathcal{Q}^\beta(\rho_{A|B_0B_1\cdots B_{N-1}})&=&\mathcal{Q}^\beta(\rho'_{A|B_0B_1\cdots B_{2^s-1}})
\leq\sum_{j=0}^{2^s-1} \mathcal{K}^{\delta\omega_H(\vec{j})}_\beta{\mathcal{Q}^\beta}(\sigma_{AB_j})
=\sum_{j=0}^{N-1} \mathcal{K}^{\delta\omega_H(\vec{j})}_\beta{\mathcal{Q}^\beta}(\rho_{AB_j}).
\end{eqnarray}

[\emph{Corollary 2}] For any multipartite state $\rho_{AB_0\cdots B_{N-1}}$, if
$k^\delta {\mathcal{Q}^\gamma_{AB_i}}\geq \sum_{l=i+1}^{N-1}{\mathcal{Q}^\gamma}_{AB_{l}}$ for $i=0, 1, \cdots, m$, and
${\mathcal{Q}^\gamma_{AB_j}}\leq k^\delta \sum_{l=j+1}^{N-1}{\mathcal{Q}^\gamma}_{AB_{l}}$ for $j=m+1,\cdots,N-2$, $0 \leq m\leq N-2$ and $N\geq 3$, we have
\begin{eqnarray}\nonumber
\mathcal{Q}^\beta_{A|B_0B_1\cdots B_{N-1}}\leq \mathcal{Q}^\beta_{AB_0}+\mathcal{K}_\beta^\delta \mathcal{Q}^\beta_{AB_1}+\cdots+\mathcal{K}^{\delta m}_\beta\mathcal{Q}^\beta_{AB_m}+\mathcal{K}^{\delta (m+2)}_\beta(\mathcal{Q}^\beta_{AB_{m+1}}+\cdots+\mathcal{Q}^\beta_{AB_{N-2}})
+\mathcal{K}^{\delta (m+1)}_\beta\mathcal{Q}^\beta_{AB_{N-1}},
\end{eqnarray}
for all $0\leq\beta\leq\gamma$, $0<k^\delta \leq 1$ and $\delta\geq 1$, where
$\mathcal{K}^\delta_\beta=\frac{(1+k^\delta)^\frac{\beta}{\gamma}-1}{k^\delta \frac{\beta}{\gamma}}$.

[\emph{Proof}] From (\ref{ab1}), we have
\begin{eqnarray}\label{pfn1}
\mathcal{Q}^{\beta}_{A|B_0B_1\cdots B_{N-1}}&&\leq {\mathcal{Q}^\beta}_{AB_0}+\mathcal{K}^\delta_\beta\left(\sum_{i=1}^{N-1}{\mathcal{Q}^\gamma}_{AB_i}\right)^\frac{\beta}{\gamma}\nonumber\\
&&\leq {\mathcal{Q}^\beta}_{AB_0}+\mathcal{K}_\beta^\delta{\mathcal{Q}^\beta}_{AB_1}
 +\left(\mathcal{K}^\delta_\beta\right)^2\left(\sum_{i=2}^{N-1}{\mathcal{Q}^\gamma}_{AB_i}\right)^\frac{\beta}{\gamma}\nonumber\\
&& \leq \cdots\nonumber\\
&&\leq {\mathcal{Q}^\beta}_{AB_0}+\mathcal{K}_\beta^\delta{\mathcal{Q}^\beta}_{AB_1}+\cdots+\mathcal{K}_\beta^{\delta m}
{\mathcal{Q}^\beta}_{AB_m}+\cdots+\mathcal{K}^{\delta(m+1)}_\beta \left(\sum_{i={m+1}}^{N-1}{\mathcal{Q}^\gamma}_{AB_i}\right)^\frac{\beta}{\gamma}.
\end{eqnarray}

Similarly, as ${\mathcal{Q}^\gamma_{AB_j}}\leq k^\delta \sum_{l=j+1}^{N-1}{\mathcal{Q}^\gamma}_{AB_{l}}$ for $j=m+1,\cdots,N-2$, we get
\begin{eqnarray}\label{pfn2}
\left(\sum_{i={m+1}}^{N-1}{\mathcal{Q}^\gamma}_{AB_i}\right)^\frac{\beta}{\gamma}
&&\leq \mathcal{K}_\beta{\mathcal{Q}^\beta_{AB_{m+1}}}+\left(\sum_{i={m+2}}^{N-1}{\mathcal{Q}^\gamma}_{AB_i}\right)^\frac{\beta}{\gamma}\nonumber\\
&&\leq \mathcal{K}^\delta_\beta\left({\mathcal{Q}^\beta}_{AB_{m+1}}+\cdots+{\mathcal{Q}^\beta}_{AB_{N-2}}\right)
+\cdots+{\mathcal{Q}^\beta}_{AB_{N-1}}.
\end{eqnarray}
Combining the inequalities (\ref{pfn1}) and (\ref{pfn2}), we prove the Corollary. $\hfill\Box$

Particularly, if $m=N-2$ in corollary 2, then one has the following corollary.

[\emph{Corollary 3}] For any multipartite state $\rho_{AB_0\cdots B_{N-1}}$, if $k^\delta \mathcal{Q}^\gamma_{AB_i}\geq\sum_{j=i+1}^{N-1} \mathcal{Q}^\gamma_{AB_j}$ for $i=0,1,\cdots N-2$, then for $0\leq\beta\leq \gamma$, $0<k^\delta \leq 1$ and $\delta\geq 1$ we have
$\mathcal{Q}^\beta_{A|B_0B_1\cdots B_{N-1}}\leq\sum_{j=0}^{N-1}\mathcal{K}^{\delta j}_\beta{\mathcal{Q}^\beta}_{AB_j}$
where $\mathcal{K}^\delta_\beta=\frac{(1+k^\delta)^\frac{\beta}{\gamma}-1}{k^{\delta \frac{\beta}{\gamma}}}$.

According to $\omega_H(\vec{j})\leq j$, we have
\begin{eqnarray}\label{pfth23}
\mathcal{Q}^\beta_{A|B_0B_1\cdots B_{N-1}}
\leq\sum_{j=0}^{N-1} \mathcal{K}^{\delta j}_\beta{\mathcal{Q}^\beta}_{AB_j}\leq\sum_{j=0}^{N-1} \mathcal{K}^{\delta\omega_H(\vec{j})}_\beta{\mathcal{Q}^\beta}_{AB_j}
\end{eqnarray}
for any $0\leq\beta\leq \gamma$. Thus, the inequality in corollary 3 is tighter than the inequality (\ref{th92}) in Theorem 2 under the conditions.

[{\it Example 2}] Let us recall the definition of Tsallis-$q$ entanglement (TE) and Tsallis-$q$ entanglement of assistance (TEoA) in \cite{ylm}. The Tsallis-$q$ entanglement $\mathcal{T}_q(|\psi\rangle_{AB})$ for any bipartite pure state $|\psi\rangle_{AB}$ is given by $\mathcal{T}_q(|\psi\rangle_{AB})=\mathcal{T}_q(\rho_A)$.
The Tsallis-$q$ entanglement of a bipartite mixed state $\rho_{AB}$ is given by
$\mathcal{T}_q(\rho_{AB})=\min\limits_{\{p_i,|\psi_i\rangle_{AB}\}}\sum_i p_i\mathcal{T}_q(|\psi\rangle_{AB})$ with the minimum taken over all possible pure state decomposition $\rho_{AB}=\sum_i p_i |\psi\rangle_{AB}\langle\psi|$.
Dual to the Tsallis-$q$ entanglement, the Tsallis-$q$ entanglement of assistance is given by \cite{ylm},
$\mathcal{T}^a_q(\rho_{AB})=\max\limits_{\{p_i,|\psi_i\rangle_{AB}\}}\sum_i p_i \mathcal{T}_q(|\psi\rangle_{AB})$, with the maximum taken over all possible pure state decomposition $\rho_{AB}=\sum_i p_i |\psi\rangle_{AB}\langle\psi|$.

In \cite{WYJ}, the authors provide a polygamy relation satisfied by the Tsallis-$q$ entanglement of assistance,
\begin{eqnarray}\label{WYJ1}
\mathcal{T}^{a\,\beta}_q(\rho_{A|B_0B_1\cdots B_{N-1}})\leq \sum_{j=0}^{N-1}\mathcal{K}_\beta^{\omega_H(\vec{j})}\mathcal{T}^{a\,\beta}_q(\rho_{AB_j}),
\end{eqnarray}
where $\mathcal{T}^{a\,\beta}_q \equiv [\mathcal{T}^a_q]^\beta$.
From Theorem 2 we obtain a tighter polygamy relation,
\begin{eqnarray}\label{WYJ2}
\mathcal{T}^{a\,\beta}_q(\rho_{A|B_0B_1\cdots B_{N-1}})\leq \sum_{j=0}^{N-1}\mathcal{K}_\beta^{\delta\omega_H(\vec{j})}\mathcal{T}^{a\,\beta}_q(\rho_{AB_j}).
\end{eqnarray}

Let us again consider the three-qubit state $|\psi\rangle_{ABC}$ in (\ref{ex1}). For $q=2$ we have $\mathcal{T}^a_2(|\psi\rangle_{A|BC})=2\lambda_0^2(\lambda_2^2+\lambda_3^2+\lambda_4^2)$, $\mathcal{T}^a_2(\rho_{AB})=2\lambda_0^2(\lambda_2^2+\lambda_4^2)$ and
$\mathcal{T}^a_2(\rho_{AC})=2\lambda_0^2(\lambda_3^2+\lambda_4^2)$.
Take $\lambda_0=\lambda_3=\frac{1}{2}$, $\lambda_2=\frac{\sqrt{2}}{2}$ and $\lambda_1=\lambda_4=0$. We get $\mathcal{T}^a_2(|\psi\rangle_{A|BC})=\frac{3}{8}$, $\mathcal{T}^a_2(\rho_{AB})=\frac{1}{4}$, $\mathcal{T}^a_2(\rho_{AC})=\frac{1}{8}$, and $z_0\equiv \mathcal{T}^{a\,\beta}_2(|\psi\rangle_{A|BC})=(\frac{3}{8})^\beta$. By choosing $k=0.8$ and $\delta=2$, we have the upper bound of (\ref{WYJ2}),
\begin{eqnarray}
z_1\equiv\mathcal{T}^{a\,\beta}_2(\rho_{AB})+\mathcal{K}_\beta^{\delta}\mathcal{T}^{a\,\beta}_2(\rho_{AC})
=(\frac{1}{4})^\beta+\frac{1.64^\beta-1}{0.64^\beta}(\frac{1}{8})^\beta.
\end{eqnarray}
The upper bound of (\ref{WYJ1}) is given by
\begin{eqnarray}\
z_2\equiv \mathcal{T}^{a\,\beta}_2(\rho_{AB})+\mathcal{K}_\beta\mathcal{T}^{a\,\beta}_2(\rho_{AC})
=(\frac{1}{4})^\beta+\frac{1.8^\beta-1}{0.8^\beta}(\frac{1}{8})^\beta.
\end{eqnarray}
Fig. 2 shows that our bound $z_1$ is indeed tighter than the bound $z_2$ from \cite{WYJ}.
\begin{figure}
  \centering\includegraphics[width=7cm]{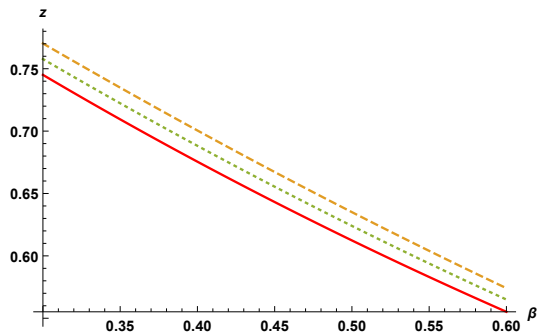}
  \caption{$z$ is the value of $\mathcal{T}^{a\,\beta}_2(|\psi\rangle_{A|BC})$. The red solid line represents the Tsallis-2 entanglement of assistance $z_0$ of $|\psi\rangle_{ABC}$ in Example 2. The green dotted line represents the upper bound $z_1$ from our result, and the yellow dashed line represents the upper bound $z_2$ from \cite{JQ}.}
\end{figure}

\section{conclusion}
The monogamy and polygamy relations give  fundamental properties of quantum correlations, which characterize the distribution of the quantum correlations in multipartite systems. We have provided a class of tighter monogamy and polygamy inequalities for general quantum correlations. Applying these results to quantum correlations such as concurrence and Tsallis-$q$ entanglement of assistance, we obtain tighter monogamy and polygamy relations that are tighter than the existing ones. Our result may highlight further studies on quantum information processing that are related to the monogamy and polygamy inequalities.

\bigskip
\noindent{\bf Acknowledgments}  This work is supported by NSFC (Grant No. 12075159), Beijing Natural Science Foundation (Z190005), Academy for Multidisciplinary Studies, Capital Normal University, the Academician Innovation Platform of Hainan Province, and Shenzhen Institute for Quantum Science and Engineering, Southern University of Science and Technology (No. SIQSE202001).


\begin{thebibliography}{99}
\bibitem{MAN} M. A. Nielsen and I. L. Chuang, \emph{Quantum Computation and Quantum Information}, (Cambridge University Press, Cambridge, England, 2000).
\bibitem{RPMK} R. Horodecki, P. Horodecki, M. Horodecki, and K. Horodecki, Quantum entanglement,
    \href{https://doi.org/10.1103/RevModPhys.81.865}{Rev. Mod. Phys. \textbf{81}, 865 (2009)}.
\bibitem{bbc} C. H. Bennett, G. Brassard, C. Cr\'epeau, R. Jozsa, A. Peres, and W. K. Wootters, Teleporting an unknown quantum state via dual classical and Einstein-Podolsky-Rosen channels, \href{https://doi.org/10.1103/PhysRevLett.70.1895}{Phys. Rev. Lett. \textbf{70}, 1895 (1993)}.
\bibitem{AGMD} A. Streltsov, G. Adesso, M. Piani, and D. Bru{\ss}, Are general quantum correlations monogamous? \href{https://doi.org/10.1103/PhysRevLett.109.050503}{Phys. Rev. Lett. \textbf{109}, 050503 (2012)}.
\bibitem{rh} R. Raussendorf and H. J. Briegel, A one-way quantum computer, \href{https://doi.org/10.1103/PhysRevLett.86.5188}{Phys. Rev. Lett. \textbf{86}, 5188 (2001)}.
\bibitem{JM} J. M. Liang, S. Q. Shen, M. Li, and S. M. Fei, Quantum algorithms for the generalized eigenvalue problem, \href{https://doi.org/10.1007/s11128-021-03370-z}{Quant. Inf. Process. 21, 23 (2022)}.
\bibitem{JML} J. M. Liang, S. Q. Shen, and M. Li, Quantum Algorithms and Circuits for Linear Equations with Infinite or No Solutions, \href{https://doi.org/10.1007/s10773-019-04151-2}{Int. J. Theor. Phys. \textbf{58}, 2632-2640 (2019)}.
\bibitem{ek} A. K. Ekert, Quantum cryptography based on Bell's theorem, \href{https://doi.org/10.1103/PhysRevLett.67.661}{Phys. Rev. Lett. \textbf{67}, 661 (1991)}.
\bibitem{LGL} G. L. Long and X. S. Liu, Theoretically efficient high-capacity quantum-key-distribution scheme, \href{https://doi.org/10.1103/PhysRevA.65.032302}{Phys. Rev. A \textbf{65}, 032302 (2002)}.
\bibitem{YJH} Y. J. Hirono, Symmetry Principle for Topologically Ordered Phases, {AAPPS Bulletin  \textbf{29}, 45-51 (2019)}.
\bibitem{XS} Y. Xiang, F. X. Sun, Q. Y. He, and Q. H. Gong, Advances in multipartite and high-dimensional Einstein-Podolsky-Rosen steering, \href{https://doi.org/10.1016/j.fmre.2020.12.003}{Fundam. Res. \textbf{1}, 99 (2021)}.
\bibitem{HWJ} W. J. Huang, W. C. Chien, C. H. Cho, C. C. Huang, T. W. Huang, and C. R. Chang, Mermin's inequalities of multiple qubits with orthogonal measurements on IBM Q 53-qubit system, \href{https://doi.org/10.1002/que2.45}{Quantum Eng. \textbf{2}, e45 (2020)}.
\bibitem{VMMP} V. Vedral, M. B. Plenio, M. A. Rippin and P. L. Knight, Quantifying entanglement, \href{https://doi.org/10.1103/PhysRevLett.78.2275} {Phys. Rev. Lett. \textbf{78}, 2275 (1997)}.
\bibitem{W} W. K. Wootters, Entanglement of formation of an arbitrary state of tow qubits, \href{https://doi.org/10.1103/PhysRevLett.80.2245} {Phys. Rev. Lett. \textbf{80}, 2245 (1998)}.
\bibitem{SW} S. Hill and W. K. Wootters, Entanglement of a pair of quantum bits, \href{https://doi.org/10.1103/PhysRevLett.78.5022}{Phys. Rev. Lett. \textbf{78}, 5022 (1997)}.
\bibitem{TMG} T. Nakano, M. Piani, and G. Adesso, Negativity of quantumness and its interpretations, \href{https://doi.org/10.1103/PhysRevA.88.012117}{Phys. Rev. A \textbf{88}, 012117 (2013)}.
\bibitem{B} B. Terhal, Is entanglement monogamous? \href{https://doi.org/10.1147/rd.481.0071}{IBM J. Res. Dev. \textbf{48}, 71 (2004)}.
\bibitem{MMV} M. Murao, M. B. Plenio, and V. Vedral, Quantum-information distribution via entanglement, \href{https://doi.org/10.1103/PhysRevA.61.032311}{Phys. Rev. A \textbf{61}, 032311 (2000)}.
\bibitem{JRomero} J. Romero, Shaping up High-dimensional Quantum Information, {AAPPS Bulletin  \textbf{29}, 2-4 (2019)}.
\bibitem{CKW} V. Coffman, J. Kundu, and W. K. Wootters, Distributed entanglement, \href{https://doi.org/10.1103/PhysRevA.61.052306}{Phys. Rev. A \textbf{61}, 052306 (2000)}.
\bibitem{TJ} T. J. Osborne and F. Verstraete, General monogamy inequality for bipartite qubit entanglement, \href{https://doi.org/10.1103/PhysRevLett.96.220503}{Phys. Rev. Lett. \textbf{96}, 220503 (2006)}.	
\bibitem{YKM} Y. K. Bai, M. Y. Ye, and Z. D. Wang, Entanglement monogamy and entanglement evolution in multipartite systems, \href{https://doi.org/10.1103/PhysRevA.80.044301}{Phys. Rev. A \textbf{80}, 044301 (2009)}.	
\bibitem{TR} T. R. de Oliveira, M. F. Cornelio, and F. F. Fanchini, Monogamy of entanglement of formation, \href{https://doi.org/10.1103/PhysRevA.89.034303}{Phys. Rev. A \textbf{89}, 034303 (2014)}.
\bibitem{agf} G. Adesso and F. Illuminati, Strong monogamy of bipartite and genuine multiparitie entanglement: the Gaussian case, \href{https://doi.org/10.1103/PhysRevLett.99.150501}{Phys. Rev. Lett. \textbf{99}, 150501 (2007)}.
\bibitem{ylm} L. M. Yang, B. Chen, S. M. Fei, Z. X. Wang, Tighter constraints of multiqubit entanglement, \href{https://doi.org/10.1088/0253-6102/71/5/545}{Commun. Theor. Phys. \textbf{71}, 545 (2019)}
\bibitem{sjin} Z. X. Jin and S. M. Fei, Superactivation of monogamy relations for nonadditive quantum correlation measures, \href{https://doi.org/10.1103/PhysRevA.99.032343}{Phys. Rev. A  \textbf{99}, 032343 (2019)}.
\bibitem{LD} D. Liu, Tighter constraints of quantum correlations among multipartite systems, \href{https://doi.org/10.1007/s10773-021-04770-8}{Int. J. Theor. Phys. \textbf{60}, 1455-1470 (2021)}.
\bibitem{kjs} J. S. Kim, Tsallis entropy and entanglement constraints in multiqubit systems, \href{https://doi.org/10.1103/PhysRevA.81.062328}{Phys. Rev. A \textbf{81}, 062328 (2010)}.
\bibitem{kjsg} J. S. Kim, Generalized entanglement constraints in multi-qubit systems in terms of Tsallis entropy, \href{https://doi.org/10.1016/j.aop.2016.07.021}{Ann. Phys. \textbf{373}, 197-206 (2016)}.
\bibitem{wmv} Y. X. Wang, L. Z. Mu, V. Vedral, and H. Fan, Entanglement R\'enyi-entropy, \href{https://doi.org/10.1103/PhysRevA.93.022324}{Phys. Rev. A \textbf{93}, 022324 (2016)}.
\bibitem{gg} G. Gour, D. A. Meyer, and B. C. Sanders, Deterministic entanglement of assistance and monogamy constraints, \href{https://doi.org/10.1103/PhysRevA.72.042329}{Phys. Rev. A \textbf{72}, 042329 (2005)}.
\bibitem{JQ} Z. X. Jin and C. F. Qiao, Monogamy and polygamy relations of multiqubit entanglement based on unified entropy, \href{https://doi.org/10.1088/1674-1056/ab6720}{Chin. Phys. B \textbf{29}, 020305 (2020)}.
\bibitem{fjin} Z. X. Jin and S. M. Fei, Polygamy relations of multipartite entanglement beyond qubits, \href{https://doi.org/10.1088/1751-8121/ab0ed9}{ J. Phys. A: Math. Theor. \textbf{52} 165303 (2019)}.
\bibitem{qjin} Z. X. Jin, S. M. Fei and C. F. Qiao, Complementary quantum correlations among multipartite systems, \href{https://doi.org/10.1007/s11128-020-2598-6}{Quant. Inf. Process. \textbf{19}, 101 (2020)}.
\bibitem{Kim} J. S. Kim, Tsallis entropy and general polygamy of multiparty quantum entanglement in arbitrary dimensions, \href{https://doi.org/10.1103/PhysRevA.94.062338}{Phys. Rev. A \textbf{94}, 062338 (2016)}.
\bibitem{ggt} N. Gisin, G. Ribordy, W. Tittel, and H. Zbinden, Quantum cryptography, \href{https://doi.org/10.1103/RevModPhys.74.145}{Rev. Mod. Phys. \textbf{74}, 145 (2002)}.
\bibitem{LYY} Y. Y. Liang, C. J. Zhu, Z. J. Zheng, Tighter monogamy constraints in multi-qubit entanglement systems, \href{https://doi.org/10.1007/s10773-020-04406-3}{Int. J. Theor. Phys. \textbf{59}, 1291-1305 (2020)}.
\bibitem{ARA} A. Kumar, R. Prabhu, A. Sen(De), and U. Sen, Effect of a large number of parties on the monogamy of quantum correlations, \href{https://doi.org/10.1103/PhysRevA.91.012341}{Phys. Rev. A \textbf{91}, 012341, (2015)}.
\bibitem{SPAU} K. Salini, R. Prabhu, A. Sen(De), and U. Sen, Monotonically increasing functions of any quantum correlation can make all multiparty states monogamous, \href{https://doi.org/10.1016/j.aop.2014.06.001}{Ann. Phys. \textbf{348}, 297-305 (2014)}.
\bibitem{WYJ} Y. J. Wu, K. Wu, Tighter weighted relations of the Tsallis-$q$ entanglement, \href{https://doi.org/10.1007/s10773-019-04297-z}{Int. J. Theor. Phys. \textbf{59}, 114-124 (2020)}.
\end{thebibliography}
\end{document}